\documentclass[prc,preprintnumbers,twocolumn]{revtex4}
\usepackage{graphicx}
\usepackage{amssymb}
\usepackage{amsmath}

\begin{document}

\title{Threshold energy for sub-barrier fusion hindrance phenomenon}
\author{V.V.Sargsyan$^{1,2}$, G.G.Adamian$^{1}$, N.V.Antonenko$^{1}$, W. Scheid$^3$, and  H.Q.Zhang$^4$
}
\affiliation{$^{1}$Joint Institute for Nuclear Research, 141980 Dubna, Russia\\
$^{2}$International Center for Advanced Studies, Yerevan State University, 0025 Yerevan, Armenia\\
$^{3}$Institut f\"ur Theoretische Physik der
Justus--Liebig--Universit\"at,
D--35392 Giessen, Germany\\
$^{4}$China Institute of Atomic Energy,  102413 Beijing,  China
}
\date{\today}

\begin{abstract}
The relationship between  the threshold energy for a deep sub-barrier
fusion hindrance phenomenon and the energy at which the regime
of interaction changes (the turning-off of the nuclear forces and friction)
in the sub-barrier capture process, is  studied within the quantum
diffusion approach. The quasielastic barrier distribution
is shown to be a useful tool to clarify whether the slope of capture cross section
changes at  sub-barrier energies.
\end{abstract}


 \maketitle


The experiments with various medium-light and heavy nuclei
have shown that the experimental slopes
of the complete fusion excitation function
keep increasing at low sub-barrier energies and may become much larger than the
predictions of standard coupled-channel calculations.
This was identified as the fusion hindrance with the threshold energy $E_s$~\cite{Jiang,Gomes,Bertulani}.
More experimental and theoretical studies of sub-barrier fusion hindrance
are required to improve our understanding of its physical reason,
which may be especially important in astrophysical fusion reactions~\cite{Zvezda}.

As shown within the quantum diffusion approach~\cite{EPJSub1,EPJSub2,EPJSub3,EPJSub4,EPJSub5},
due to a change of the regime of interaction (the turning-off of the nuclear forces and friction)
at deep sub-barrier energies, the curve related to the  capture cross section
as a function of bombarding energy has smaller slope.
In the present paper we try to demonstrate the relationship between  the threshold energy $E_s$
for a deep
sub-barrier
fusion hindrance phenomenon and the energy $E_{ch}$ at which the
regime of interaction  changes in the sub-barrier capture process.

In the quantum diffusion approach
the capture of  nuclei is treated in terms
of a single collective variable: the relative distance  between
the colliding nuclei. The  neutron transfer and nuclear deformation effects
are taken into consideration through the dependence of the nucleus-nucleus potential
on the isotopic compositions,  deformations and orientations of interacting nuclei.
Our approach takes into consideration the fluctuation and dissipation effects in
collisions of heavy ions which model the coupling with various channels
(for example, coupling of the relative motion with low-lying collective modes
such as dynamical quadrupole and octupole modes of target and projectile~\cite{Ayik333}).
We have to mention that many quantum-mechanical and non-Markovian effects~\cite{Hofman,Ayik,Hupin} accompanying
the passage through the potential barrier are taken into consideration in our
formalism~\cite{EPJSub1,EPJSub2,EPJSub3,EPJSub4,EPJSub5}.
The details of  used formalism are presented in our previous  articles~\cite{EPJSub1,EPJSub2}.
With this approach many heavy-ion capture
reactions at energies above and well below the Coulomb barrier have been
successfully described.

Within the quantum diffusion model~\cite{EPJSub1,EPJSub2,EPJSub3,EPJSub4,EPJSub5}
 the nuclear forces start to play a role
at relative distance $R_{int}=R_b+1.1$ fm
($R_b$ is the position of the Coulomb barrier at given angular momentum and orientations of the interacting nuclei)
where the nucleon density of colliding nuclei approximately reaches
10\% of saturation density.
If the colliding nuclei approach the distance $R_{int}$ between their centers, the
nuclear forces start to act in addition to the Coulomb interaction. Thus, at $R<R_{int}$
the relative motion may be more coupled with other degrees of freedom. At $R>R_{int}$
the relative motion is almost independent of the internal degrees of freedom.
Depending on whether the value of external turning point
$R_{ex}$ is larger or smaller than interaction radius $R_{int}$,
the impact of coupling with other degrees of freedom upon the barrier passage
seems to be different.
So, two regimes of interaction at sub-barrier energies differ by the action
of nuclear forces and, respectively, of nuclear friction.
Due to the switching-off the nuclear interaction at external turning point $R_{ex}$,
the cross sections falls with the smaller rate at a deep sub-barrier energies.
\begin{figure}[h]
\vspace*{-0.cm}
\centering
\includegraphics[angle=0, width=0.9\columnwidth]{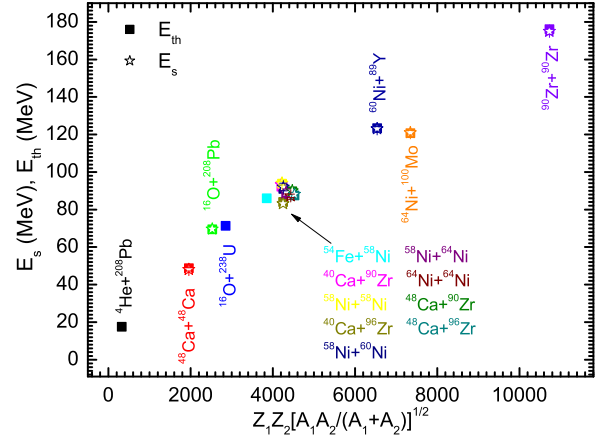}
\vspace*{-0.2cm}
\caption{The experimental threshold energy $E_{s}$
for a deep
sub-barrier
fusion hindrance phenomenon~\protect\cite{Jiang} and the calculated energy $E_{ch}$ at which  the
regime of interaction changes in the indicated  sub-barrier capture reactions as
a function of $Z_1Z_2[A_1A_2/(A_1+A_2)]^{1/2}$.
}
\label{1_fig}
\end{figure}

As seen in Fig.~1,
for the reactions  $^{4}$He + $^{208}$Pb,
$^{58}$Ni + $^{54}$Fe, $^{48}$Ca + $^{48}$Ca,$^{90,96}$Zr,
$^{40}$Ca + $^{90,96}$Zr, $^{58}$Ni + $^{58,60,64}$Ni, $^{60}$Ni + $^{89}$Y, $^{64}$Ni + $^{64}$Ni,$^{100}$Mo,
$^{90}$Zr+$^{90}$Zr, and $^{16}$O + $^{208}$Pb,$^{238}$U,
there is a good agreement between the threshold energy $E_s$
for a deep
sub-barrier
fusion hindrance phenomenon and the energy $E_{ch}$ at which  the
regime of interaction changes in the sub-barrier capture process.
The values $E_s$ and  $E_{ch}$  almost coincide and linearly increase with $Z_1Z_2[A_1A_2/(A_1+A_2)]^{1/2}$.

and the capture cross section is the sum of the fusion and quasifission cross sections,
from the comparison of calculated capture cross sections and measured
fusion cross sections one can extract the hindrance factor and the threshold incident energy for a deep
sub-barrier fusion hindrance phenomenon.
The small fusion cross section at energies well below the Coulomb barrier
may indicate that the quasifission channel is preferable and
the system goes to this channel after the capture~\cite{EPJSub1,EPJSub2,EPJSub3,EPJSub4,EPJSub5}.
So, the observed hindrance factor may be understood in term of quasifission.
At deep sub-barrier energies,
the quasifission event  corresponds to the formation
of a nuclear-molecular state or dinuclear system
with small excitation energy that separates
(in the competition with  the  compound nucleus formation process)
by the quantum tunneling through the Coulomb barrier
in a binary event with mass and charge  close to the colliding nuclei.
In this sense the quasifission is the general phenomenon
which takes place in the reactions with the
massive~\cite{GSI,Volkov,nasha,Avaz}, and medium-mass  nuclei~\cite{EPJSub2}.
\begin{figure}
\vspace*{-0.cm}
\centering
\includegraphics[angle=0, width=0.85\columnwidth]{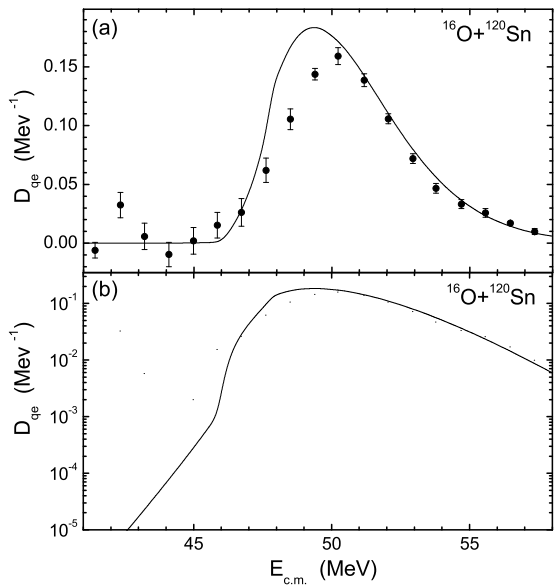}
\vspace*{-0.2cm}
\caption{The calculated (solid line) $D_{qe}(E_{\rm c.m.})=dP_{cap}(E_{\rm c.m.},J=0)/dE_{\rm c.m.}$
for the $^{16}$O + $^{120}$Sn reaction.
The experimental data (symbols) are from Ref.~\protect\cite{Sinha}.
The values of $D_{qe}(E_{\rm c.m.})$ are shown in the linear (a) and logarithmic (b) scales.
}
\label{2_fig}
\end{figure}

Since
the quasielastic measurements are usually not as complex as the capture (fusion)
measurements, and  they are well suited to survey the decreasing
rate of fall of the sub-barrier capture cross section.
There is a direct relationship between the capture
 and the quasielastic scattering processes, because any loss from
the quasielastic channel contributes directly to the capture (the conservation of the reaction flux):
$$P_{qe}(E_{\rm c.m.},J)+P_{cap}(E_{\rm c.m.},J)=1$$
and
$$dP_{cap}/dE_{\rm c.m.}=-dP_{qe}/dE_{\rm c.m.},$$
where
$P_{qe}$  is the reflection probability and
$P_{cap}$ is the capture (transmission) probability.
The quasielastic scattering is the sum of
 elastic, inelastic, and transfer processes.
The reflection probability
$$P_{qe}(E_{\rm c.m.},J=0)=d\sigma_{qe}/d\sigma_{Ru}$$
for
angular momentum $J=0$ is given by the ratio of
the quasielastic differential cross section  and
Rutherford differential cross section at 180 degrees~\cite{Timmers,Timmers2,Zhang,Sonzogni,Sinha,Piasecki}.
\begin{figure}
\vspace*{-0.cm}
\centering
\includegraphics[angle=0, width=0.85\columnwidth]{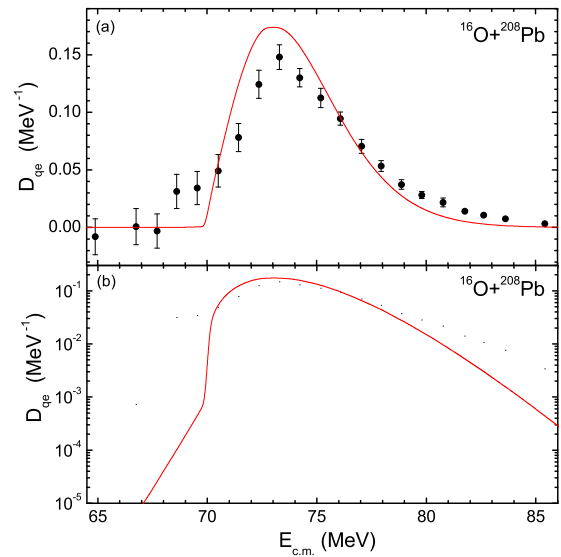}
\vspace*{-0.2cm}
\caption{The same as in Fig.~2 but
for the  $^{16}$O + $^{208}$Pb reaction.
 The experimental data (symbols) are from Ref.~\protect\cite{Timmers2}.
}
\label{3_fig}
\end{figure}
\begin{figure}
\vspace*{-0.cm}
\centering
\includegraphics[angle=0, width=0.85\columnwidth]{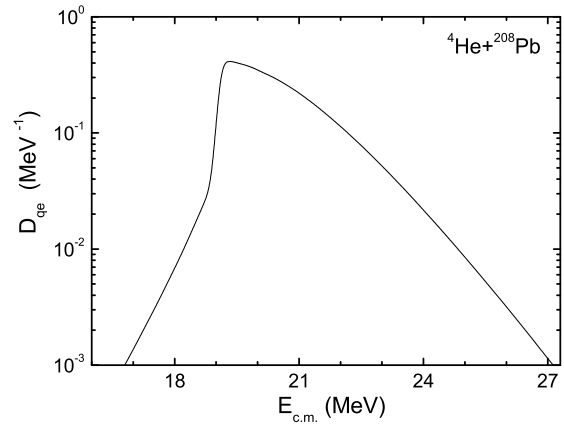}
\vspace*{-0.2cm}
\caption{The calculated (solid line)
$D_{qe}(E_{\rm c.m.})=dP_{cap}(E_{\rm c.m.},J=0)/dE_{\rm c.m.}$
for the $^{4}$He + $^{208}$Pb reaction.
}
\label{4_fig}
\end{figure}
 The barrier distribution is extracted by taking
the first derivative of the $P_{qe}$ with respect to $E_{\rm c.m.}$, that is,
$$D_{qe}(E_{\rm c.m.})=-dP_{qe}(E_{\rm c.m.},J=0)/dE_{\rm c.m.}=$$
$$=dP_{cap}(E_{\rm c.m.},J=0)/dE_{\rm c.m.}.$$
Thus, one can  observe the change of the fall rate of $P_{cap}(E_{c.m.},J=0)$ at sub-barrier energies
by measuring the barrier distribution $D_{qe}$.
By employing the quantum diffusion approach and
calculating $dP_{cap}(E_{\rm c.m.},J=0)/dE_{\rm c.m.}$,
one can obtain $D_{qe}(E_{\rm c.m.})$.
In addition to the mean peak position of the  $D_{qe}$ around
the barrier height, we predict the sharp change of the slope of  $D_{qe}$  below
the threshold energy  because of a change of the
regime of interaction  in the sub-barrier capture process (Figs.~2--4).
The effect seems to
be more pronounced in the collisions of spherical nuclei (Figs.~3 and 4).
The collisions of deformed nuclei occurs at various mutual
orientations on which the value of $R_{int}$ depends.
Thus, the deformation and neutron transfer  effects
can smear out this  effect.
 The reactions
$^{4}$He,$^{16}$O + $^{A}$Sn,$^{144}$Sm,$^{208}$Pb and $^{48,40}$Ca$,^{36}$S+$^{90}$Zr with the spherical nuclei
are preferable for the experimental study of $D_{qe}(E_{\rm c.m.})$.


In conclusions, employing the quantum diffusion approach, we demonstrated
the relationship between  the threshold energy
for a deep sub-barrier
fusion hindrance phenomenon and the energy at which the
regime of interaction changes in the sub-barrier capture process.
We predicted the sharp change of the slope of
 the quasielastic barrier distribution
below the threshold energy.
This is expected to be the experimental indication
of a  change of the regime of interaction  in the sub-barrier capture. One concludes that
the quasielastic technique could be an  important tool in capture (fusion) research.

This work was supported by DFG, NSFC, and RFBR.
The IN2P3(France)-JINR(Dubna) and Polish - JINR(Dubna)
Cooperation Programmes are gratefully acknowledged.\\


\begin{thebibliography}{99}
\bibitem{Jiang}
C.L.~Jiang~{\it et al.}, Phys. Rev. Lett. {\bf 89}, 052701 (2002);
C.L.~Jiang~{\it et al.}, Phys. Rev. C {\bf 71}, 044613 (2005);
C.L.~Jiang, B.B.~Back, H.~Esbensen, R.V.F.~Janssens, and K.E.~Rehm, Phys. Rev. C {\bf 73}, 014613 (2006);
H.~Esbensen and C.L.~Jiang, Phys. Rev. C {\bf 79}, 064619 (2009).
\bibitem{Gomes} L.F.~Canto, P.R.S.~Gomes, R.~Donangelo, and M.S.~Hussein,
 Phys. Rep. {\bf 424}, (2006) 1.
\bibitem{Bertulani} C.A.~Bertulani, EPJ Web Conf. {\bf 17}, 15001 (2011).

\bibitem{Zvezda}        K.~Langanke and  C.A.~Barnes, Adv.Nucl.Phys. {\bf 22},  (1996)  173;
A.~Aprahamian, K.~Langanke, and M.~Wiescher,   Prog.Part.Nucl.Phys. {\bf 54}, (2005) 535.
%
\bibitem{EPJSub1}        V.V.~Sargsyan, G.G.~Adamian, N.V.~Antonenko,  W.~Scheid, and H.Q.~Zhang,
Eur. Phys. J. A {\bf 47}, 38 (2011);
J. of Phys.: Conf. Ser. {\bf 282}, 012001 (2011);  EPJ Web Conf. {\bf 17}, 04003 (2011).
\bibitem{EPJSub2}        V.V.~Sargsyan, G.G.~Adamian, N.V.~Antonenko,  W.~Scheid, and H.Q.~Zhang,
Phys. Phys. C {\bf 84}, 064614 (2011); Phys. Rev. C {\bf 85}, 024616 (2012); Phys. Rev. C {\bf 85}, 069903 (2012).
\bibitem{EPJSub3}        V.V.~Sargsyan, G.G.~Adamian, N.V.~Antonenko, and W.~Scheid, Eur. Phys. J. A {\bf 45}, 125 (2010).
\bibitem{EPJSub4}        V.V.~Sargsyan, G.G.~Adamian, N.V.~Antonenko,  W.~Scheid, C.J.~Lin, and H.Q.~Zhang,
Phys. Phys. C {\bf 85},  017603 (2012);
Phys. Phys. C {\bf 85},  037602 (2012).
\bibitem{EPJSub5} R.A.~Kuzyakin, V.V.~Sargsyan, G.G.~Adamian, N.V.~Antonenko, E.E.~Saperstein, and S.V.~Tolokonnikov,
Phys. Rev. C {\bf 85}, 034612 (2012).
\bibitem{Ayik333} S.~Ayik, B.~Yilmaz, and D.~Lacroix,  Phys. Rev. C {\bf 81}, 034605 (2010).
%
\bibitem{Hofman}        H.~Hofmann, Phys. Rep.  {\bf 284}, 137 (1997);
                        C.~Rummel and H.~Hofmann, Nucl. Phys. A {\bf 727}, 24 (2003).
\bibitem{Ayik} N.~Takigawa, S.~Ayik, K.~Washiyama, and S.~Kimura, Phys. Rev. C {\bf 69}, 054605 (2004);
S.~Ayik, B.~Yilmaz, A. Gokalp, O. Yilmaz, and N.~Takigawa, Phys. Rev. C {\bf 71}, 054611 (2005).
\bibitem{Hupin} G.~Hupin and D.~Lacroix,  Phys. Rev. C {\bf 81}, 014609 (2010).

\bibitem{GSI}           J.G.~Keller {\it et al.},  Nucl. Phys. {\bf A452},  173 (1986).
\bibitem{Volkov}       V.V.~Volkov, Particles and Nuclei, {\bf 35},  797 (2004).
\bibitem{nasha}     G.G.~Adamian, N.V.~Antonenko, and~W.Scheid, Phys. Rev. C {\bf 68},  034601  (2003);
{\it  Lecture Notes in Physics} {\bf 848}, {\it Clusters in Nuclei},
Vol. 2, ed. by C.~Beck (Springer-Verlag, Berlin, 2012) p.~165.
\bibitem{Avaz}           G.~Giardina~{\it et al.}, Nucl. Phys.  {\bf A671},  165  (2000);
                        A.~Nasirov {\it et al.}, Nucl. Phys.  {\bf A759},  342 (2005);
Z.-Q.~Feng, G.-M.~Jin, J.-Q.~Li, and W. Scheid,
Phys. Rev. C {\bf 76}, 044606 (2007);~H.Q.~Zhang, C.L.~Zhang, C.J.~Lin, Z.H.~Liu, F.~Yang, A.K.~Nasirov, G.~Mandaglio,
M.~Manganaro, and G.~Giardina, Phys. Rev. C {\bf 81}, 034611 (2010).

\bibitem{Timmers}     H.~Timmers, J.R.~Leigh, M.~Dasgupta, D.J.~Hinde,
R.C.~Lemmon, J.C.~Mein, C.R.~Morton, J.O.~Newton, and N.~Rowley,
Nucl. Phys.  {\bf A584}, 190 (1995).
\bibitem{Timmers2}    H.~Timmers, Ph. D. thesis, Australian National University (1996).
\bibitem{Zhang}    H.Q.~Zhang, F.~Yang, C.~Lin, Z.~Liu, F.~Yang, and Y.~Hu, Phys. Rev. C {\bf 57}, R1047 (1998);
Inter. workshop on {\it Nuclear reactions and beyond, Wold Scientific (2000), p.95}; F.~Yang, C.J.~Lin, X.K.~Wu, H.Q.~Zhang,
 C.L.~Zhang, P.Zhou, and Z.H.~Liu, Phys. Rev. C {\bf 77}, 014601 (2008).
\bibitem{Sonzogni}    A.A.~Sonzogni, J.D.~Bierman, M.P.~Kelly, J.P.~Lestone, J.F.~Liang, and
R.~Vandenbosch, Phys. Rev. C {\bf 57}, 722 (1998).
\bibitem{Sinha}    S.~Sinha, M.R.~Pahlavani, R.~Varma, R.K.~Choudhury, B.K.~Nayak, and A.~Saxena,
Phys. Rev. C {\bf 64}, 024607 (2001).
\bibitem{Piasecki}    E.~Piasecki {\it et al.}, Phys. Rev. C {\bf 65}, 054611 (2002);
E.~Piasecki {\it et al.}, Phys. Rev. C {\bf 80}, 054613 (2009);
E.~Piasecki {\it et al.}, Phys. Rev. C {\bf 85}, 054604 (2012); {\it ibid} {\bf 85}, 054608 (2012).

\end{thebibliography}
\end{document}